\documentstyle[floats,aps,prl,epsfig,twocolumn,amstex]{revtex}

\begin{document}
\input amssym.def
\input amssym
\bibliographystyle{/usr/share/texmf/tex/latex/revtex/prsty}
\draft
\wideabs{
\title{The Three- and Four-Nucleon Systems from Chiral Effective Field
  Theory}

\author{
E. Epelbaum$^{\ddagger}$, H. Kamada$^{\ddagger, *}$,
A. Nogga$^*$, H. Wita\l a$^{\dagger}$, W. Gl\"ockle$^*$, Ulf-G. Mei\ss ner$^{\ddagger}$
}

\address{$^\ddagger $ Forschungszentrum J\"ulich, Institut f\"ur 
Kernphysik (Theorie), 52425 J\"ulich, Germany}
\address{$^*$
Institut f\"ur  Theoretische Physik II,
         Ruhr-Universit\"at Bochum, 44780 Bochum, Germany}
\address{$^\dagger $ Institute of Physics, Jagellonian University, 
PL-30059 Cracow, Poland} 

\date{\today}
\maketitle

\begin{abstract}
Recently developed chiral nucleon--nucleon (NN) forces 
at next-to-leading order (NLO) that describe NN phase
shifts up to about 100 MeV fairly well have been applied to 3N and 4N systems.
Faddeev-Yakubovsky equations have been solved rigorously. The chiral NLO forces depend on a
momentum cut-off $\Lambda$ lying between 540-600 MeV/c. The resulting 3N and 4N binding energies are
in the same range as found using standard NN potentials. In additon, low--energy
3N scattering observables are very well reproduced like for standard NN forces. Surprisingly,
the long standing $A_y$--puzzle is resolved at NLO. The cut-off
dependence of the scattering observables is rather mild.
\end{abstract}
\pacs{PACS number(s): 21.45.+v,21.30.-x,27.10.+h,25.10.+s}
}
\narrowtext

Effective field theory (EFT) has become a standard tool in modern physics and
is applied to a large variety of systems. It can also be used to construct
nuclear forces in a systematic and controlled manner. The spontaneously and 
explicitely broken chiral symmetry of QCD can be implemented in the EFT formulated 
in terms of the asymptotically observed Goldstone boson (pion) and
matter (nucleon) fields. In the purely pionic and the pion-nucleon systems,
there is an expansion parameter which  is a typical external momentum (or the
quark masses) divided
by a hadronic mass scale of the order of the $\rho$ meson or the
nucleon mass. Any S-matrix element or transition current can  be
systematically expanded in terms of this small parameter based on a systematic
power counting. In systems with more than one nucleon, an additional
non-perturbative resummation is mandatory to deal with the shallow nuclear
bound states. This idea was put forward by Weinberg~\cite{Weinberg} and
taken up by van Kolck and collaborators~\cite{Kolck,Ordonez} 
in the construction of two- (NN) and  three-nucleon (3N) forces. 
One basically constructs a potential
based on the power counting and calculates bound and scattering states by use of
a properly regularized Lippmann-Schwinger or Schr\"odinger equation. One
outstanding result was that 3N forces (3NF) vanish to leading
order~\cite{Weinberg}. Other
groups also investigated low energy properties of NN systems along
these lines~\cite{park,munich}.
A different counting scheme was proposed by Kaplan et al.~\cite{Kaplan} (KSW) 
working directly with the scattering amplitudes instead with an effective
Hamiltonian like it is the case  along the lines proposed by Weinberg. 
Another important feature which distinguishes the KSW approach from
Weinberg's is the perturbative treatment of the one-pion
exchange. Independent of these differences, such type of framework for the 
first time offers the possibility of calculating nuclear
forces directly from fundamental principles and has a direct link to the
chiral properties of QCD. Furthermore, this approach is firmly based
on quantum field theory and avoids ill-defined concepts like
meson-nucleon form factors.

In \cite{Epelbaum1} we have taken up Weinberg's
idea and constructed a NN and 3N potential
based on the most general  effective chiral pion-nucleon Lagrangian using the method of
unitary transformations. In this method the field theoretical pion-nucleon Hamiltonian is
decoupled such that an effective purely nucleonic Hamiltonian consistent with a power
counting scheme arises. Previous results were obtained in 
time-ordered perturbation theory, which lead to energy-dependent
nuclear forces. In our formalism, we arrive at hermitian energy-independent
nuclear forces  which we consider to be a major
advantage with respect to applications to 3N  and 4N systems, the issue of this
letter. In\cite{Weinberg,Ordonez,Epelbaum1,Epelbaum2} NN forces have been developed at 
leading, next-to-leading and next-to-next-to-leading orders, LO, NLO and 
NNLO, respectively. At LO the potential is represented by the ordinary one-pion exchange 
(with a point-like coupling)
as well as two contact interactions without derivatives. At NLO one includes the leading 
chiral two-pion exchange as well as all possible contact interactions with two derivatives,
whereas  at NNLO we have additional two-pion exchange with low-energy constants (LECs)
determined from pion-nucleon scattering~\cite{paul}. The forces are properly
renormalized and contain nine parameters related to those four-fermion contact terms. The
one- and two-pion exchange pieces are parameter free. The nine LEC´s have been uniquely
fixed to low energy NN phase shifts in the s- and p-waves. The parameter free predictions
for higher energies and partial waves and also deuteron properties are in
general  rather  good. It was also observed that the NNLO predictions are
better than the ones based on the NLO potential, as expected in a systematic EFT. 

%

\begin{table}[tb]
\caption{\label{table1}
Neutron-proton phase shifts in our approach (upper row) compared to the Nijmegen
PSA (middle row) and the CD-Bonn potential (lower row).} 
\vskip 0.3 true cm
\begin{tabular} {c@{\hspace{5mm}}|r@{\hspace{5mm}}|r@{\hspace{5mm}}|r@{\hspace{5mm}}|r@{\hspace{5mm}}}
&\multicolumn{1}{c}{ 1 MeV } &\multicolumn{1}{c}{ 5 MeV } 
&\multicolumn{1}{c}{  10 MeV } & \multicolumn{1}{c}{ 20 MeV  }\\
\hline
 & 62.044  & 63.869     & 60.28 & 53.76  \\
$^1S_0$& 62.078 & 63.645     & 59.97 & 53.56 \\
 & 62.069 & 63.627     & 59.96 & 53.57 \\ \hline
 & 147.695  & 118.308     & 102.87 &  86.33 \\
$^3S_1$& 147.748 & 118.175     & 102.60 & 86.09 \\
 & 147.747  & 118.178     & 102.61 & 86.12 \\ \hline
 & 0.107  & 0.679     & 1.17 &  1.65 \\
$\epsilon_1$& 0.105 & 0.674     & 1.16 & 1.66 \\
 & 0.105  & 0.672     & 1.16 & 1.66 \\ \hline
 & -0.005  & -0.181     & -0.67 &  -2.07 \\
$^3D_1$& -0.005 & -0.184     & -0.68 & -2.06 \\
 & -0.005  & -0.183     & -0.68 & -2.05 \\ \hline
 & -0.187  & -1.493     & -3.08 & -5.54 \\
$^1P_1$& -0.189 & -1.503     & -3.08 & -5.47 \\
 & -0.187  & -1.487     & -3.04 & -5.40 \\ \hline
 & 0.187  & 1.676     & 3.73 &  7.06 \\
$^3P_0$& 0.177 & 1.608     & 3.62 & 6.92 \\
 & 0.180  & 1.626     & 3.65 & 6.95 \\ \hline
 & -0.117  & -0.994     & -2.16 & -4.18 \\
$^3P_1$& -0.108 & -0.932     & -2.05 & -4.01 \\
 & -0.108  & -0.937     & -2.06 & -4.03 \\ \hline
 & 0.020  & 0.237     & 0.70 &  2.05 \\
$^3P_2$& 0.022 & 0.255     & 0.72 & 1.84 \\
 & 0.022  & 0.251     & 0.71 & 1.84 \\  
\end{tabular}
\end{table}

The natural question arises now, whether the NN forces based on chiral perturbation theory
($\chi$PT)
will be also successful in describing 3N and 4N low-energy observables. To that aim we
solve the Faddeev-Yakubovsky equations rigorously for 3N and 4N
systems \cite{Nogga,Gloeckle4N,Kamada} and determine binding
energies and various scattering observables. To the best of our knowledge this is the first
time that $\chi$PT has been practically applied to nuclear systems beyond $A=2$
within the Hamiltonian approach. 

In this first application we restrict ourselves to the NLO NN potential. In a forthcoming
article we shall go on to NNLO and include also 3NF´s, which occur at that order the first
time. The NLO results presented here are therefore parameter free and can
serve as a good testing ground for the usefulness of the approach.
Of course, some aspects of the 3N system have already been studied in
nuclear EFT~\cite{Bedaque,Griess}, but not as direct extensions of the
NN system as done here.

In order to use the chiral NN forces in the NN Lippmann-Schwinger equation one has
to introduce a momentum regulator $\Lambda$. 
We remark that this regularization on the level of the scattering equation is
completely different from standard methods which are applied to individual
diagrams.  Here we use the smooth regulator, its precise  form is given in~\cite{Epelbaum2}.
In order to investigate the cut-off dependence of 3N and 4N observables we
have generated several NN potentials corresponding to different cut-offs
between $\Lambda$ =540 and 600 MeV/c.  They were all fitted to the $^1S_0$,
$^3S_1$-$^3D_1$, $^1P_1$ and $^3P_{0,1,2}$ NN phase shifts 
up to $E_{\rm lab} = 100\,$MeV (for the potential parameters contact E.E.).
In~\cite{Epelbaum2} we had already demonstrated that going to
higher order in the EFT reduces the cut-off dependence and allows to choose 
larger values for the  cut-offs,
as expected from general arguments~\cite{lepage}.
The resulting phase shifts for NLO are compared to the Nijmegen phase shift
analysis \cite{Nijm} and the ones obtained from the CD-Bonn potential~\cite{Machleidt}  in
Table~\ref{table1}. The agreement is fairly well and we know from \cite{Epelbaum2}  
that one has to go to  NNLO to improve systematically on that. Also, we
restrict ourselves to the isospin symmetric case. Thus we do not take into account various 
charge independence and charge symmetry breaking effects like e.g. the pion mass difference, 
which are known to be significant at very low energies.
Such effects can also be included systematically in our EFT~\cite{Markus}.

\begin{table}[tb]
    \caption{\label{tab:bind}   Theoretical $^3$H and $^4$He binding energies for
     different cut-offs $\Lambda$  
     compared to CD-Bonn predictions and to the experimental $^3$H binding energy and the Coulomb
     corrected $^4$He binding energy.  
    The kinetic energies and $S'$, $P$ and $D$ 
    state probabilities for $^4$He are also shown. }
\vskip 0.3 true cm
    \begin{tabular}[t]{r|r|rr|rrr}
Potential & $E_T$ [MeV] &  $E_{^4\rm{He}}$ &   $T$ [MeV]  
                & $S'$ [\%] & $P$ [\%] & $D$ [\%] \\
\hline
NLO, 540     &  -8.284 & -28.03   & 65.2   &  0.62   & 0.08  &  6.00   \\
NLO, 560     &  -8.091 & -26.91   & 68.2   &  0.68   & 0.09  &  6.41   \\
NLO, 580     &  -7.847 & -25.55   & 72.2   &  0.76   & 0.10  &  6.84   \\
NLO, 600     &  -7.546 & -23.96   & 77.7   &  0.86   & 0.11  &  7.30   \\
CD-Bonn &  -8.012 & -27.05   & 77.6   &  0.48   & 0.22  &  10.72  \\ 
\hline	                    
exp     &  -8.48  & -29.00   & ---    &  ---    & ---   &  ---    \\
     \end{tabular}
\end{table}

Let us regard  3N and 4N binding energies now. 
In addition to the relative motion in the NN 
subsystem there occur 
relative motions of the third and fourth nucleon now. In the spirit of a low momentum theory
one has to expect that those additional relative momenta should be also small. This turns
out to be true as  the numerics tells us. 
For standard NN potentials, which are defined for all momenta up to infinity, one introduces
a momentum cut--off for the numerical integrations. Since the pertinent integrals are convergent,
the results are cut--off independent.
Using the chiral NN potential we could
reduce the cut-offs in those relative momenta from their typical
values 8 fm$^{-1}$ arising with standard potentials
down to 4 fm$^{-1}$  without changing the result. 
In the Faddeev-Yakubovsky equations 
there is no mechanism which could introduce high
momenta if the  NN subsystem momenta are small to start
with. We find for the fully
converged solutions the 3N and 4N binding energies as given in Table
\ref{tab:bind}. 
The ranges are compatible with what is found using realistic
potentials \cite{Nogga}. Note that the
NN forces were included up to the total 
NN angular momentum of $j_{\rm max}$=6.

\begin{figure}[tb]
  \begin{center} 
    \epsfxsize=8.5cm
    \epsffile{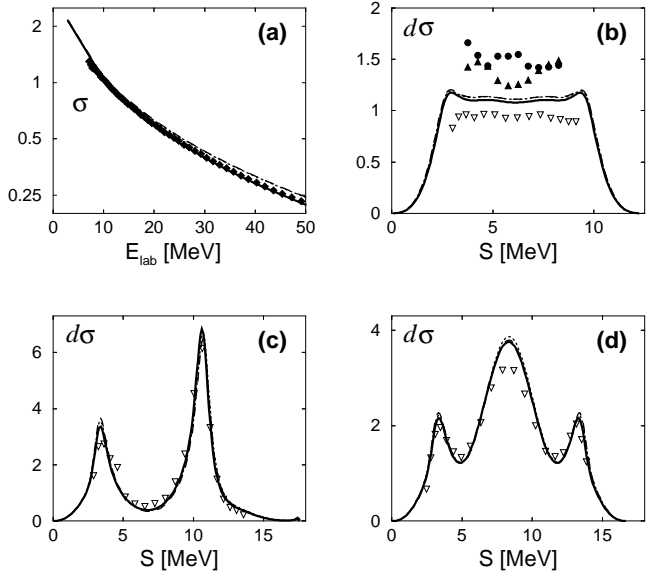}
    \caption{
    (a): The {\it nd}- total cross section (in barn) for the chiral forces ($\Lambda = 540$~MeV/c,
    dotted curve; $\Lambda = 600$~MeV/c, dashed curve), and CD-Bonn (thick solid curve).
    (b), (c) and (d): Chiral NN
    force and CD-Bonn predictions for {\it nd} break-up cross section  
    $d \sigma \equiv { d^5\sigma \over d \Omega_1 d \Omega_2 dS}$ 
    [${\rm mb \over sr^2 MeV}$] at $E_{\rm lab}=$13 MeV along the kinematical locus S. 
    The various break-up configurations are chosen 
    according to Figs.~42, 39 and 35 in \protect \cite{Gloeckle}, respectively. 
    {\it pd} data are ($\scriptstyle \triangledown$) 
    \protect\cite{Rauprich2}; {\it nd} data are ($\scriptstyle \blacktriangle$) 
    \protect\cite{Strate}, ($\bullet$) \protect\cite{Setze}, 
   ($\scriptstyle \blacklozenge$) \protect\cite{Abfalterer}.
\label{fig:stot} 
 }
  \end{center}
\end{figure}

\begin{figure}[tb]
  \begin{center} 
    \epsfxsize=8.5cm
    \epsffile{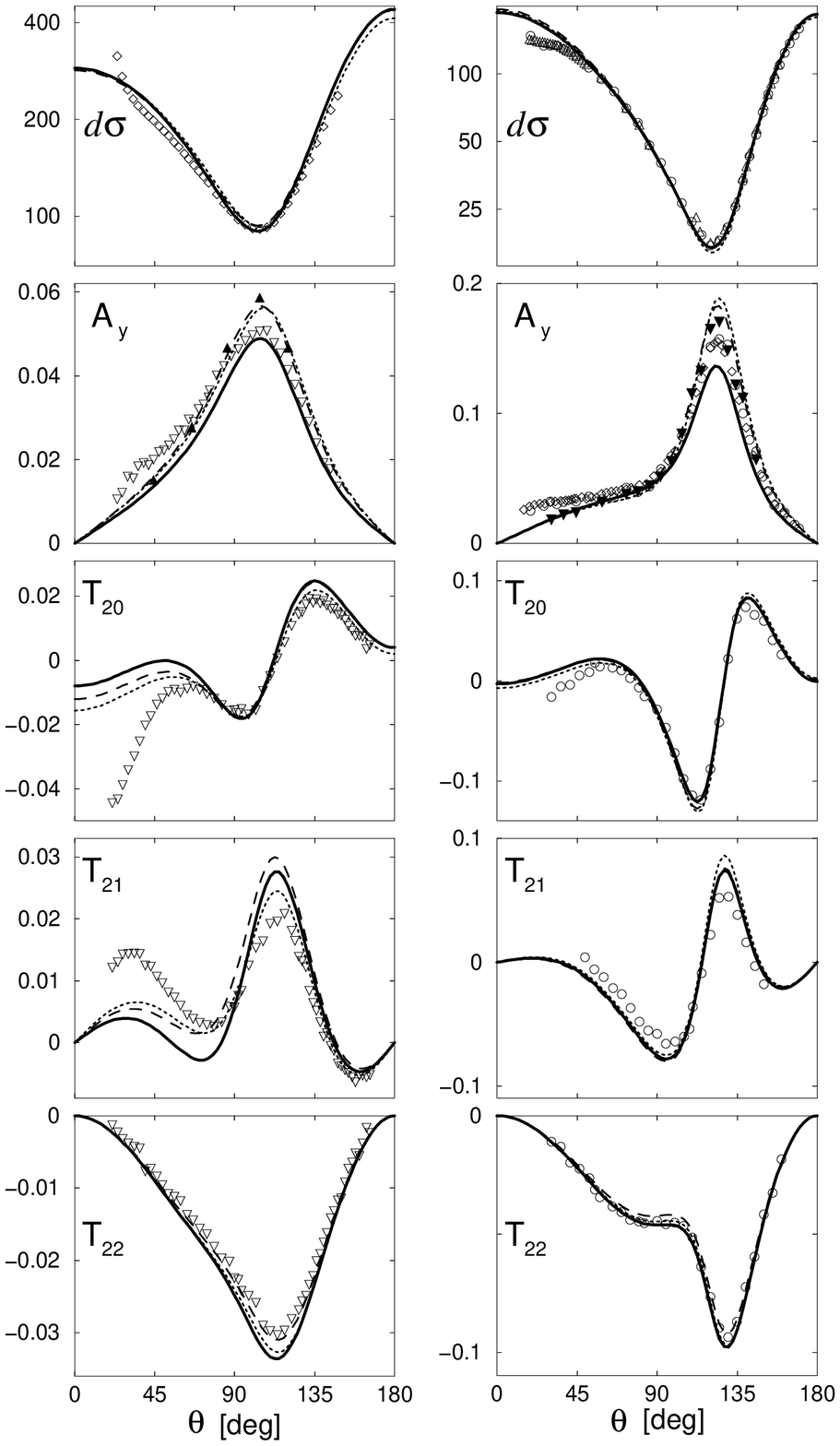}
\caption{
   {\it nd} elastic scattering observables at $E_n=3$MeV (left
    column) and $E_n=10$MeV (right
    column). {\it pd} data are ($\diamond$) 
   \protect \cite{Sagara}, ($\scriptstyle \triangledown$) \protect\cite{Shimizu},
    ($\scriptstyle \vartriangle$) \protect\cite{Rauprich1}, ($\circ$) \protect\cite{Sperisen}. 
    {\it nd} data are ($\scriptstyle \blacktriangle$) 
    \protect\cite{McAninch}, ($\scriptstyle \blacktriangledown$) 
    \protect\cite{Howell}. For remaining notations see Fig.~\protect\ref{fig:stot}.
\label{fig:elastic}
} 
    
  \end{center}
\end{figure}


%

Also 3N scattering can be solved rigorously nowadays\cite{Gloeckle} 
and we show in Figs.~\ref{fig:stot}, \ref{fig:elastic}  a small
selection out of the great wealth of observables in comparison to data and the theoretical predictions
of CD-Bonn. Three
energies, one below the {\it nd} break-up threshold and two above are chosen.
Note further that we do not indicate error bars for the data since in most cases they will be 
not distinguishable on this scale. 
In all cases
we show the predictions of the chiral NN potentials for the cut-offs $\Lambda=540$ and 
$\Lambda=600$~MeV/c and compare it to the 
prediction of one of the most modern, so--called realistic NN potentials, the
CD-Bonn\cite{Machleidt}. As the simplest observable we show first the
{\it nd} total  cross section in Fig.~\ref{fig:stot}(a). The three theoretical curves overlap 
completely at very low energies and then the chiral predictions start to deviate somewhat from 
the data as  expected for our EFT at NLO.  
 
In case of the 3N break-up we selected in Fig.~\ref{fig:stot}(b,c,d) a few
often investigated configurations, the space-star, a final 
state interaction peak configuration and a quasi--free scattering (QFS) configuration, respectively.  
We find very good agreement of the chiral NN force predictions with the
one from CD-Bonn. 
In case of  QFS (d) and space-star (b) some of the discrepancies are expected to be
caused by Coulomb force effects not included in the theory. The upper group of data in (b)
are {\it nd} data and the disagreement with the theory  presents a well known puzzle at the 
moment \cite{Howell1}.

\begin{figure}[tb]
  \begin{center} 
   \epsfxsize=8.5cm
    \epsffile{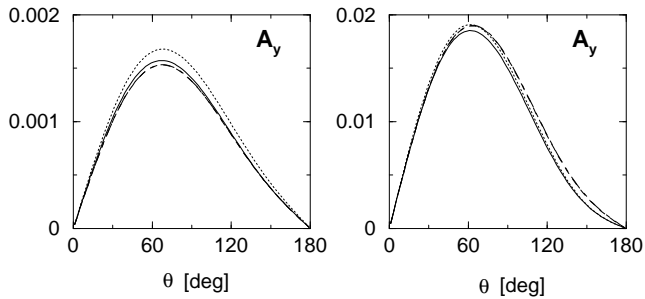}
    \caption{np analyzing powers $A_y$ at 3 (left panel) and 12 MeV (right panel) for the chiral forces ($\Lambda = 540$~MeV/c,
dashed curve; $\Lambda = 600$~MeV/c, dot-dashed curve), PSA (solid curve) and CD-Bonn (dotted curve).}
    \label{fig:nnanal}
  \end{center}
\end{figure}

For elastic
{\it nd} scattering we display in Fig.~\ref{fig:elastic} the angular
distributions, the nucleon analyzing power $A_y$ and the tensor analyzing powers $T_{2k}$.
Except for $A_y$ there are no {\it nd} data for those energies.
The discrepancies between data and theory for $T_{2k}$ and for the differential cross section
can be traced back to {\it pp} Coulomb force effects    
\cite{Kievsky}. 
Thus except for $A_y$ the agreement of CD-Bonn (thick solid curve) with the data is rather
good, which is a well known fact and is just given for the sake of 
orientation. The dotted and dashed 
curves are the chiral NN force predictions for $\Lambda = 540$ and $\Lambda
= 600$~MeV/c, respectively. 
The $d \sigma$ and $T_{2k}$ agree rather well with the CD-Bonn
result and thus with 
the data. We consider this to be an important result, demonstrating that the 
chiral NN forces are very well suited to also describe low-energy 3N scattering observables
rather quantitatively. On top of that, 
surprisingly for us, the chiral force predictions are now
significantly higher in the 
maxima of $A_y$ than for CD-Bonn and break the long standing situation that
all standard realistic NN 
forces up to now underpredict the maxima by about 30~\%. This is called 
the $A_y$--puzzle\cite{Gloeckle}. We are now in fact rather close to the 
experimental {\it nd} values.
Since we restrict ourselves to NLO we can not expect a final
answer from the 
point of view of chiral dynamics but this result for $A_y$ 
is very interesting. In that context 
it is important to note that on a 2N level
the chiral potential predictions agree well with the predictions based in the 
Nijmegen PSA as is shown in Fig.~\ref{fig:nnanal}. There we also include the CD-Bonn predictions. 
This agreement is especially important for the {\it np} analyzing power $A_y$, since it
is rather sensitive to the $^3P_j$ NN phase shifts, which influence also strongly the 3N $A_y$.

These very first results using chiral NN forces in 3N and 4N systems
are very promising. Since 
we restricted ourselves to NLO we could not expect a very good
description of the data, 
since on-shell properties are not perfectly well fitted. But the
results show that these 
effective chiral forces are very well suited to describe also  low
energy properties of 
nuclear systems beyond $A=2$. 
They agree rather well with standard nuclear force predictions as
exemplified with CD-Bonn 
and most importantly they break the stagnation in the $A_y$ puzzle.
Our result provides a counter example to the suggestion \cite{Hueb} that NN forces alone can
not describe 2N and 3N $A_y$'s at the same time and 3NF's should cure the 3N $A_y$ puzzle.
Examples for such trials can be found in \cite{Kievsky,Canton}.

It will be very interesting to perform the
next step and use the 
NNLO NN forces, which is a systematic improvement. On this level also
3N forces have to be 
incorporated the first time, which in  $\chi$PT  are defined
consistently with the NN force. It should be
mentioned further that due to the underlying Lagrangian the coupling of external probes (photons for
instance) is well defined and exchange currents consistent  to the nuclear forces can be
generated. Also the hybrid approaches \cite{Weinberg2,Bernard} 
can be avoided and wavefunctions based on the chiral
dynamics can be used directly instead of wavefunctions generated by standard NN forces. This
appears advisable since in this study for NLO we noticed that kinetic energies in 3N and 4N bound
states are smaller and two-nucleon correlation functions are smoother than for standard
NN forces, indicating a change of the wave functions.

This work was supported by the Deutsche Forschungsgemeinschaft (A.N.) and by the Polish Committee for Scientific Research under Grant No. 2P03B02818 (H.W.).
The numerical calculations have been performed on the CRAY T90 and T3E of the 
John von Neumann Institute for Computing 
in J\"ulich, Germany.

\end{document}